\documentclass[letterpaper, 12pt]{article}\usepackage[]{graphicx}\usepackage[]{color}
\makeatletter
\def\maxwidth{ %
  \ifdim\Gin@nat@width>\linewidth
    \linewidth
  \else
    \Gin@nat@width
  \fi
}
\makeatother

\definecolor{fgcolor}{rgb}{0.345, 0.345, 0.345}

\usepackage{framed}
\makeatletter
 {\par\unskip\endMakeFramed%
 \at@end@of@kframe}
\makeatother

\definecolor{shadecolor}{rgb}{.97, .97, .97}
\definecolor{messagecolor}{rgb}{0, 0, 0}
\definecolor{warningcolor}{rgb}{1, 0, 1}
\definecolor{errorcolor}{rgb}{1, 0, 0}

\usepackage{alltt}
\usepackage{a4wide}
\usepackage[hidelinks]{hyperref}
\usepackage{setspace}
\usepackage{graphicx}
\usepackage{amsmath}
\usepackage{amssymb}
\usepackage{float}
\usepackage{caption}
\usepackage{subcaption}
\usepackage{afterpage}
\usepackage{adjustbox} 
\usepackage{microtype} 
\usepackage{enumitem} 

\usepackage[authoryear,comma,longnamesfirst,sectionbib]{natbib} 
\usepackage{mathptmx}
\IfFileExists{upquote.sty}{\usepackage{upquote}}{}

\title{Bayesian survival analysis of batsmen in Test cricket}
\author{Oliver G. Stevenson$^{1*}$ and Brendon J. Brewer$^1$\\
\vspace{0cm}\\
$^1$\normalsize{Department of Statistics, The University of Auckland}\\
\normalsize{Private Bag 92019, Auckland 1142, New Zealand}\\
\vspace{0cm}\\
$^*$\normalsize{Email: {\tt o.g.stevenson@gmail.com}}
}

\date{}

\begin{document}

\maketitle

\section*{Abstract}
Cricketing knowledge tells us batting is more difficult early in a player's
innings but becomes easier as a player familiarizes themselves with the
conditions.
In this paper, we develop a Bayesian survival analysis method to predict the
Test Match batting abilities for international cricketers.
The model is applied in two stages, firstly to individual players, allowing us
to quantify players' initial and equilibrium batting abilities, and the rate
of transition between the two.
This is followed by implementing the model using a hierarchical structure,
providing us with more general inference concerning a selected group of
opening batsmen from New Zealand.
The results indicate most players begin their innings playing with between only
a quarter and half of their potential batting ability.
Using the hierarchical structure we are able to make predictions for the batting
abilities of the next opening batsman to debut for New Zealand.
Additionally, we compare and identify players who excel in the role of opening
the batting, which has practical implications in terms of batting order and
team selection policy.
\\
\\
\noindent \textbf{Key words:} Bayesian survival analysis, hierarchical
modelling, cricket

\section{Introduction}

Since the inception of statistical record-keeping in cricket, a player's
batting ability has primarily been recognised using a single number,
their batting average.
However, in cricketing circles it is common knowledge that a player will
not begin an innings batting to the best of their ability. 
Rather, it takes time to adjust both physically and mentally to the specific 
match conditions. 
This process is nicknamed `getting your eye in'. 
External factors such as the weather and the state of the pitch are rarely 
the same in any two matches and can take time to get used to. 
Additionally, batsmen will often arrive at the crease with the match poised 
in a different situation to their previous innings, requiring a different 
mental approach. 
Subsequently, batsmen are regularly seen to be dismissed early in their 
innings while still familiarizing themselves with the conditions. 
This suggests that a constant-hazard model, whereby the probability of a 
batsman being dismissed on their current score (called the \textit{hazard}) 
remains constant regardless of their score, is not ideal for predicting when 
a batsman will get out. 

It would be of practical use to both coaches and players to have a more 
flexible method of quantifying how well a batsman is performing at 
any given stage of their innings.
Identifying players' batting weaknesses and improving team selection can be 
aided by tools that estimate measures such as (1) how well a batsman performs 
when they first arrive at the crease, (2) how much better they perform once 
they have `got their eye in' and (3) how long it takes them to accomplish this.
In this paper, we propose a Bayesian parametric model to identify how an
individual batsman's ability changes over the course of an innings.

Given the data-rich nature of the sport, numerous studies have used metrics such
as batting average to optimize player and team performance.
Cricketing data has been used to fine-tune both playing strategies
\citep{clarke1988, clarke1999, preston2000, davis2015}, and decision making
\citep{clarke2003, swartz2006, norman2010} during a match.
Yet, surprisingly few studies have focused on developing new player performance
measures to better explain batting ability than the humble batting average.

Pre-computing, \cite{elderton1945} provided empirical evidence to support
the claim that a batsman's scores could be modelled using a geometric
progression.
However, the geometric assumption does not necessarily hold for all players
\citep{kimber1993}, namely due to its difficulty in fitting the inflated
number of scores of 0 appearing in many players' career records.
To account for this, \cite{bracewell2009} proposed to model player batting
scores using the `Ducks `n' runs' distribution, using a beta distribution to
model scores of zero, and a geometric to describe the distribution of
non-zero scores.

Rather than model batting scores, \cite{kimber1993} used nonparametric models
to derive a player's hazard at a given score, estimating dismissal 
probabilities as a batsman's innings progresses.
Methods for estimating the hazard function for discrete and ordinal data have
long-existed in survival analysis \citep{mccullagh1980, allison1982}, and have
applications across a wide range of disciplines.
However, the present case may be considered unusual in the context of discrete
hazard functions, given the large number of ordered, discrete points (i.e.
number of runs scored) \citep{agresti2011}.
Estimating the hazard function allows us to observe how a player's dismissal
probability (and therefore, their batting ability) varies over the course of
their innings.
While \cite{kimber1993} found batsmen were more likely to get out early
in their innings, due to the sparsity of data at higher scores these
estimates quickly become unreliable and the estimated hazard function jumps
erratically between scores.
\cite{cai2002} address this issue using a parametric smoother on the hazard
function, however given the underlying function is still a nonparametric 
estimator the problem of data sparsity still remains an issue and continues 
to distort the hazard function at higher scores.

As an alternative, \cite{brewer2008} proposed a Bayesian parametric model to 
estimate a player's current batting ability (via the hazard function)
given their score, using a single change-point model.
This allows for a smooth transition in the hazard between a batsman's 
`initial' and `eye in' states, rather than the sudden jumps seen in 
\cite{kimber1993} and to an extent \cite{cai2002}.
Based on our knowledge of cricket, it is fair to assume that batsmen are 
more susceptible early in their innings and tend to perform better as they 
score more runs.
As such, this model allows us to place appropriate priors on the parameters 
of our model to reflect our cricketing knowledge.

Bayesian stochastic methods have also been used to measure batting
performance \citep{koulis2014, damodaran2006}.
\cite{koulis2014} propose a model for evaluating performance based on player
form, however this only allows for innings to innings comparisons in terms
of batting ability, rather than comparisons \textit{during} an innings.
On the other hand, \cite{damodaran2006} provides a method which does allow for
within-innings comparisons, but lacks a natural cricketing interpretation.
Various other performance metrics have been proposed, however have been in
relation to limited overs cricket
\citep{lemmer2004, lemmer2011, damodaran2006, koulis2014}.

In this paper we focus exclusively on Test and first-class cricket, as limited
overs cricket introduces a number of complications \citep{davis2015}.
We propose an alternative Bayesian model to \cite{brewer2008} for inferring
a batsman's hazard function from their career batting record.
For now, using a nonparametric approach within a Bayesian context would
afford the hazard function far too much freedom and would result in poorly
constrained inferences if applied to individual players \citep{brewer2008}.
While the model is simple, it provides a foundation to which we can add a
more complex structure to in the future.
Additionally, in this paper we use the model as part of a hierarchical
inference, allowing us to make generalised statements about a wider group
of players, rather than being restricted to analysing a single player at a
time.
We can therefore quantify how player abilities differ, taking into account the
fact that the information about any particular player is limited by the
finite number of innings they have played.

\section{Model Specification}

The derivation of the model likelihood follows the method detailed in 
\cite{brewer2008}.
In cricket, a player bats and continues to score runs until: (1) 
he is dismissed, (2) every other player in his team is dismissed,
(3) his team's innings is concluded via a declaration or (4) the match 
ends.
Consider the score $X \in \{0, 1, 2, 3, ...\}$ that a batsman scores in a 
particular innings. 
Define the hazard function, $H(x) \in [0, 1]$, as the probability the batsman 
scores $x$ ($P(X = x)$), given they are currently on score $x$;
i.e., the probability the batsman scores no more runs

\begin{equation}
  H(x) = P(X = x | X \geq x) = \frac{P(X = x, X \geq x)}{P(X \geq x)} = \frac{P(X = x)}{P(X \geq x)}.
\label{hx}
\end{equation}

Throughout this section, all probabilities and distributions are conditional 
on some set of parameters, $\theta$, which will determine the form of $H(x)$ 
and therefore $P(X = x)$. 
We proceed by defining $G(x) = P(X \geq x)$ as the `backwards' cumulative 
distribution. 
Using this definition, Equation \ref{hx} can be written as a difference 
equation for $G(x)$:
 
\begin{equation}
\begin{split}
  G(x) & = P(X \geq x) \\
  G(x) & = P(X = x) + P(X \geq x + 1) \\
  G(x) & = H(x) G(x) + G(x + 1) \\
  G(x + 1) & = G(x) - H(x) G(x) \\
  G(x + 1) & = G(x) [1 - H(x)].
\end{split}
\end{equation}

With the initial condition $G(0) = 1$ and an assumed functional 
form for $H(x)$, we can calculate $G(x)$ for $x > 0$:

\begin{equation}
  G(x) = \prod_{a = 0}^{x - 1} [1 - H(a)].
\end{equation}

This is the probability of scoring one run, times the probability of scoring 
two runs given that you scored one run, etc., up to the probability of scoring 
$x$ runs given that you scored $x - 1$ runs. 
Therefore, the probability distribution for $X$ is given by the probability 
of surviving up until score $x$, then being dismissed:

\begin{equation}
  P(X = x) = H(x) \prod_{a = 0}^{x - 1} \left[ 1 - H(a) \right].
\end{equation}

Which is the probability distribution for the score in a single innings, given 
a model of $H$.
When we infer the parameters $\theta$ from data, this expression provides
the likelihood function.
For multiple innings we assume conditional independence, and for not-out 
innings we use $P(X\geq x)$ as the likelihood, rather than $P(X = x)$.
This assumes that for not-out scores, the batsman would have gone on to score
some unobserved score, conditional on their current score and their assumed
hazard function.
If we considered these unobserved scores as additional unknown parameters and
marginalized them out, we would achieve the same results but at higher
computational cost.
Thus, if $I$ is the total number of innings and $N$ is the number of not-out 
scores, the probability distribution for a set of conditionally independent 
scores $\{x_i\}_{i = 1}^{I - N}$ and not-out scores $\{y_i\}_{i = 1}^N$ is 

\begin{equation}
  p(\{x\}, \{y\}) = \prod_{i = 1}^{I - N} \Big(H(x_i) \prod_{a = 0}^{x_i - 1} [1 - H(a)] \Big) \times \prod_{i = 1}^N \Big(\prod_{a = 0}^{y_i - 1} [1 - H(a)] \Big).
  \label{lik}
\end{equation}

When data $\{x, y\}$ are fixed and known, Equation \ref{lik} above gives the 
likelihood for any proposed model of $H(x; \theta)$, the hazard function.
The log-likelihood is

\begin{equation}
  \textup{log}\left[L(\theta)\right] = \sum_{i = 1}^{I - N} \textup{log} \ H(x_i) + \sum_{i = 1}^{I - N} \sum_{a = 0}^{x_i - 1} \textup{log} [1 - H(a)] + \sum_{i = 1}^{N} \sum_{a = 0}^{y_i - 1} \textup{log}[1 - H(a)]
\label{loglik}
\end{equation}

where $\theta$ is the set of parameters controlling the form of $H(x)$.

\section{Hazard Function}

The parameterization of the hazard function $H(x)$ will influence how well we can 
fit the data, as well as what we can learn from doing so.
In order to accurately reflect our belief that batsmen are more susceptible 
to being dismissed early in their innings, the hazard function should be 
higher for low values of $x$ (i.e. low scores) and decrease as $x$ increases, 
as the batsman gets used to the match conditions.

If $H(x) = h$, is a constant value $h$, then the sampling distribution
$P(X = x)$ is simply a geometric distribution with expectation
$\mu = \frac{1}{h} - 1$, similar to the approach used by \cite{elderton1945}.
If we think in terms of $\mu$, it makes sense to parameterize the hazard
function in terms of an `effective batting average', $\mu(x)$, which evolves
with score as a batsman gets their eye in. 
This allows us to think of playing ability in terms of batting averages 
rather than dismissal probabilities, which has a more natural 
interpretation to the everyday cricketer. 
We can obtain $H(x)$ from $\mu(x)$ as follows:

\begin{equation}
  H(x) = \frac{1}{\mu(x) + 1}.
\end{equation}

Therefore the hazard function, $H(x)$, relies on our parameterization of a 
player's effective batting average, $\mu(x)$. 
It is reasonable to consider a batsman beginning their innings playing with 
some initial playing ability $\mu(0) = \mu_1$, which increases with the 
number of runs scored until a peak playing ability $\mu_2$ is reached. 
\citet{brewer2008} used a sigmoidal model for the transition from 
$\mu_1$ to $\mu_2$.
However, it is both simpler and probably more realistic to adopt a functional 
form for $\mu(x)$ where the transition from $\mu_1$ to $\mu_2$ necessarily 
begins immediately, and where $\mu(0) = \mu_1$ by definition.
Therefore we adopt an exponential model where $\mu(x)$ begins at $\mu_1$ and 
approaches $\mu_2$ as follows:

\begin{equation}
  \mu(x; \mu_1, \mu_2, L) = \mu_2 + (\mu_1 - \mu_2)\exp\left(-\frac{x}{L}\right).
  \label{mux}
\end{equation}

Our model contains just three parameters: $\mu_1$ and $\mu_2$, the initial 
and equilibrium batting abilities of the player, and $L$, the timescale of 
the transition between these states. 
Formally, $L$ is the $e$-folding time and can be understood by analogy with 
a `half-life', signifying the number of runs to be scored for 63\% of the
transition between $\mu_1$ and $\mu_2$ to take place.
The major change between the present model and that of \citet{brewer2008} is
that we use just a single parameter, $L$, to describe the transition between
the two effective average parameters, and that $\mu_1$ has a natural
interpretation since it equals $\mu(0)$.

Since we do not expect a batsman's ability to decrease once arriving at the
crease, we impose the constraint $\mu_1 \leq \mu_2$.
However, it is worth noting there are various instances during a Test
where this assumption may be violated.
Batting often becomes more difficult due to a deterioration in physical
conditions such as the pitch or light.
The introduction of a new bowler or new type of bowler (e.g. a spin bowler)
may also disrupt the flow of a batsman's innings, especially when the change
coincides with the bowling side opting to take the new ball after 80
(or more) overs.
Additionally, batsmen are likely to take some time re-adjusting to the
conditions after a lengthy break in play, particularly when resuming their
innings at the start of a new day's play.
However, data on these possible confounders is difficult to obtain and it is
not clear that including them in the model and then integrating over their
related parameters would lead to a large difference from our current approach
of ignoring these effects because we do not have the relevant data.

We also do not expect the transition between the two batting states to be
any larger than the player's `eye in' effective batting average, so we
also restrict the value of $L$ to be less than or equal to $\mu_2$.
To implement these constraints, we performed the inference by
re-parameterizing from $(\mu_1, \mu_2, L)$ to $(C, \mu_2, D)$  such that 
$\mu_1 = C\mu_2$ and $L = D\mu_2$, where C and D are restricted to the interval
[0, 1].
In terms of the three parameters $(C, \mu_2, D)$, the effective average model
is

\begin{equation}
  \mu(x; C, \mu_2, D) = \mu_2 + \mu_2(C - 1)\exp \left(-\frac{x}{D\mu_2}\right).
\label{effectiveaverage}
\end{equation}

Therefore the hazard function takes the form

\begin{equation}
  H(x) = \frac{1}{\mu_2 + \mu_2(C - 1)\exp \left(-\frac{x}{D\mu_2}\right) + 1}.
\end{equation}

\begin{figure}[H]
\centering
  \includegraphics[width = 0.85\linewidth]{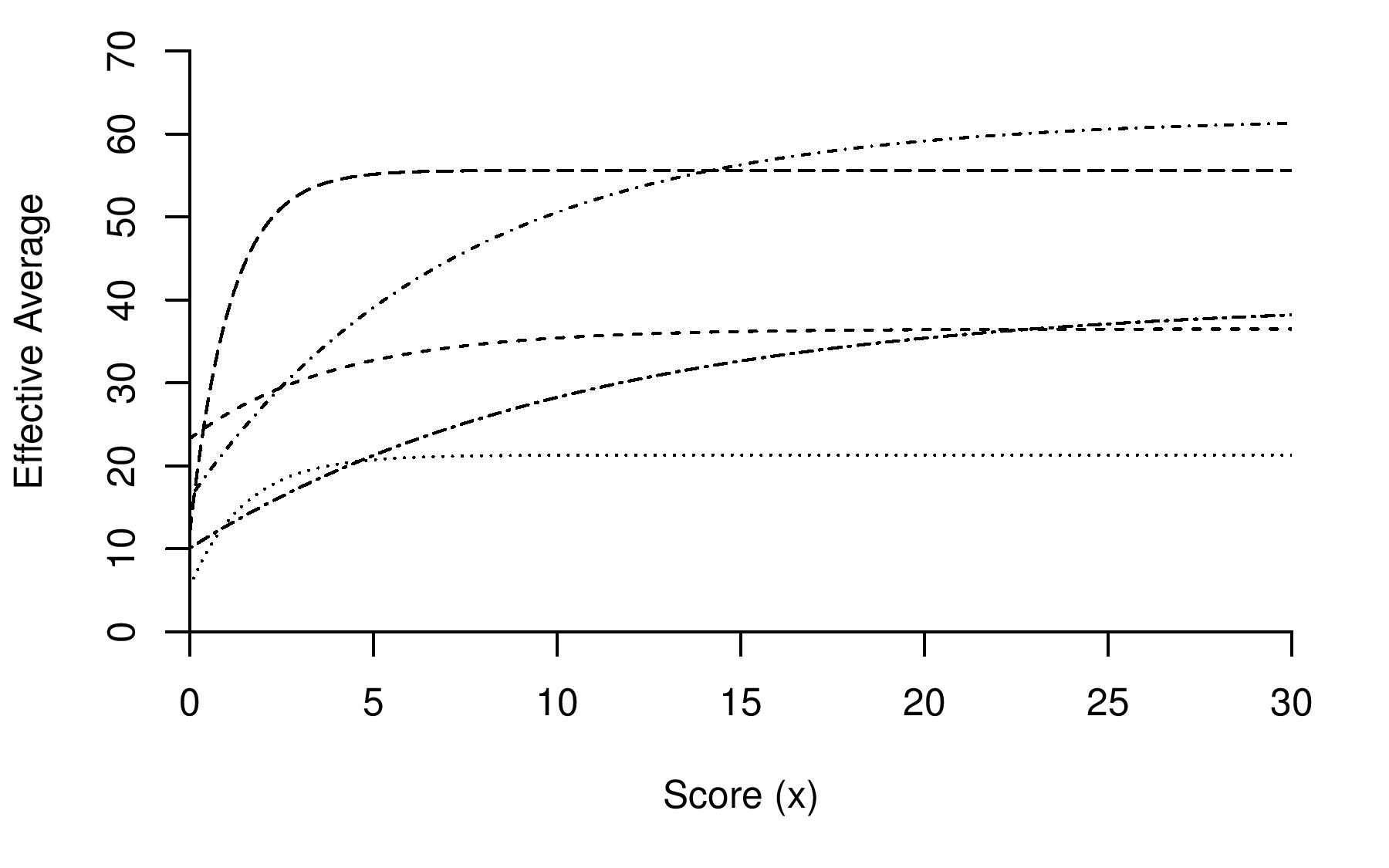}
\caption{Examples of various plausible effective average functions $\mu(x)$, 
ranging from small to large differences between the initial and equilibrium 
effective averages $\mu_1$ and $\mu_2$, with both fast and slow transition 
timescales $L$.}
\label{hxexample}
\end{figure}

See Figure \ref{hxexample} for some examples of possible effective average 
functions $\mu(x)$ allowed by this model.

\section{Analysis of Individual Players} \label{CommonPrior}

The first stage of the analysis involved evaluating individual player data, 
using fixed priors for the parameters $C$, $\mu_2$ and $D$ of each player. 
This allows us to reconstruct the joint posterior distributions for $\mu_1$, 
$\mu_2$ and $L$ for each player. 

The individual player analysis simply requires us to specify priors on 
parameters $C$, $\mu_2$ and $D$. 
All parameters are non-negative and $C$ and $D$ lie between 0 and 1. 
For $\mu_2$, we selected a prior that loosely coincided with our cricketing
knowledge.
A lognormal(25, 0.75$^2$) prior was chosen, signifying a prior median `eye in'
batting average of 25, with a width (standard deviation of 
$\textup{log}(\mu_2)$) of 0.75.
The lognormal distribution was preferred as it is a natural and well-known
distribution for modelling uncertainty about a positive quantity whose
uncertainty spans an order of magnitude or so.
This prior implies an expected number of runs per wicket of 33 when batsmen
have their eye in which seems reasonable in the context of Test
cricket.
The width of 0.75 implies a conservatively wide uncertainty.
The prior 68\% and 95\% credible intervals for $\mu_2$ are $[11.81, 52.93]$
and $[5.75, 108.7]$ respectively.

Selecting a prior which considers a wider range of $\mu_2$ values is
ill-advised, as it would allow the model to fit very high `eye in' batting
abilities for a player with a small sample of high scoring innings.
In reality, it is highly improbable that any Test player will have an
effective average greater than 100 at any stage of their innings, except for
the great Donald Bradman, whose cricketing feats are unlikely to be seen again.

The joint prior for $C$ and $D$ was chosen to be independent of the prior on
$\mu_2$ and $C$ and $D$ were chosen to be independent from each other.
As both $C$ and $D$ are restricted to the interval [0, 1], we used beta(1, 2)
and beta(1, 5) priors respectively to emphasize the lower end of the interval.
These priors represent mean initial batting abilities and $e$-folding times
that are one-third and one-sixth of a player's `eye in' effective average
respectively, and allow for a range of plausible hazard functions
(see Figure 1).
The overall Bayesian model specification for analyzing an individual player 
is therefore

\begin{align}
  \mu_2 &\sim \textup{Lognormal}(25, 0.75^2) \\
  C &\sim \textup{Beta}(1, 2) \\
  D &\sim \textup{Beta}(1, 5) \\
  \textnormal{log-likelihood} &\sim \textnormal{Equation~\ref{loglik}}
\label{commonpriors}
\end{align}

The joint posterior distribution for $\mu_2$, $C$ and $D$ is proportional to 
the prior times the likelihood function.
We can then sample from the joint posterior distributions to make inferences
about an individual player's initial playing ability, `eye in' playing
ability and the abruptness of the transition between these states.

\subsection{Data}

The data were obtained from Statsguru, the cricket statistics database on
the Cricinfo website\footnote{http://www.espncricinfo.com/}, using the R
package \texttt{cricketr} \citep{cricketr}.
We used Test Match data as the model assumptions are more likely to 
be sufficiently realistic.
Players have more time to bat in Tests and therefore scores are more 
likely to reflect a player's true batting nature, rather than the match 
situation.
To assess the performance of the individual player model, we analyzed 
the same players as in Brewer's (2008) original study.
This dataset consists of an arbitrary mixture of retired batsmen, all-rounders
and a bowler, each of whom enjoyed a long Test career during the 1990s and
2000s.

To perform the computation, we used a Julia implementation
\footnote{https://github.com/eggplantbren/NestedSampling.jl} of the nested
sampling (NS) algorithm \citep{skilling2006} which uses Metropolis-Hastings
updates.
This gives us the posterior distributions of parameters $\mu_1$,
$\mu_2$ and $L$, for each player, as well as the marginal likelihood.

For each player, we used 1000 NS particles and 1000 MCMC steps per NS
iteration.
As the model only contains three parameters, simpler MCMC schemes (or even
simple Monte Carlo or importance sampling) would also work here.
Advantages of NS include being able to deal with high dimensional and
multimodal problems, which may arise as we add more parameters to our model.
As such, we used nested sampling from the beginning because it will allow us to
continue using the same method on more complex models in the future, and carry
out model selection trivially.

\subsection{Results}

\subsubsection{Marginal Posterior Distributions}
\begin{figure}[h]
\centering
  \includegraphics[width = 0.85\linewidth]{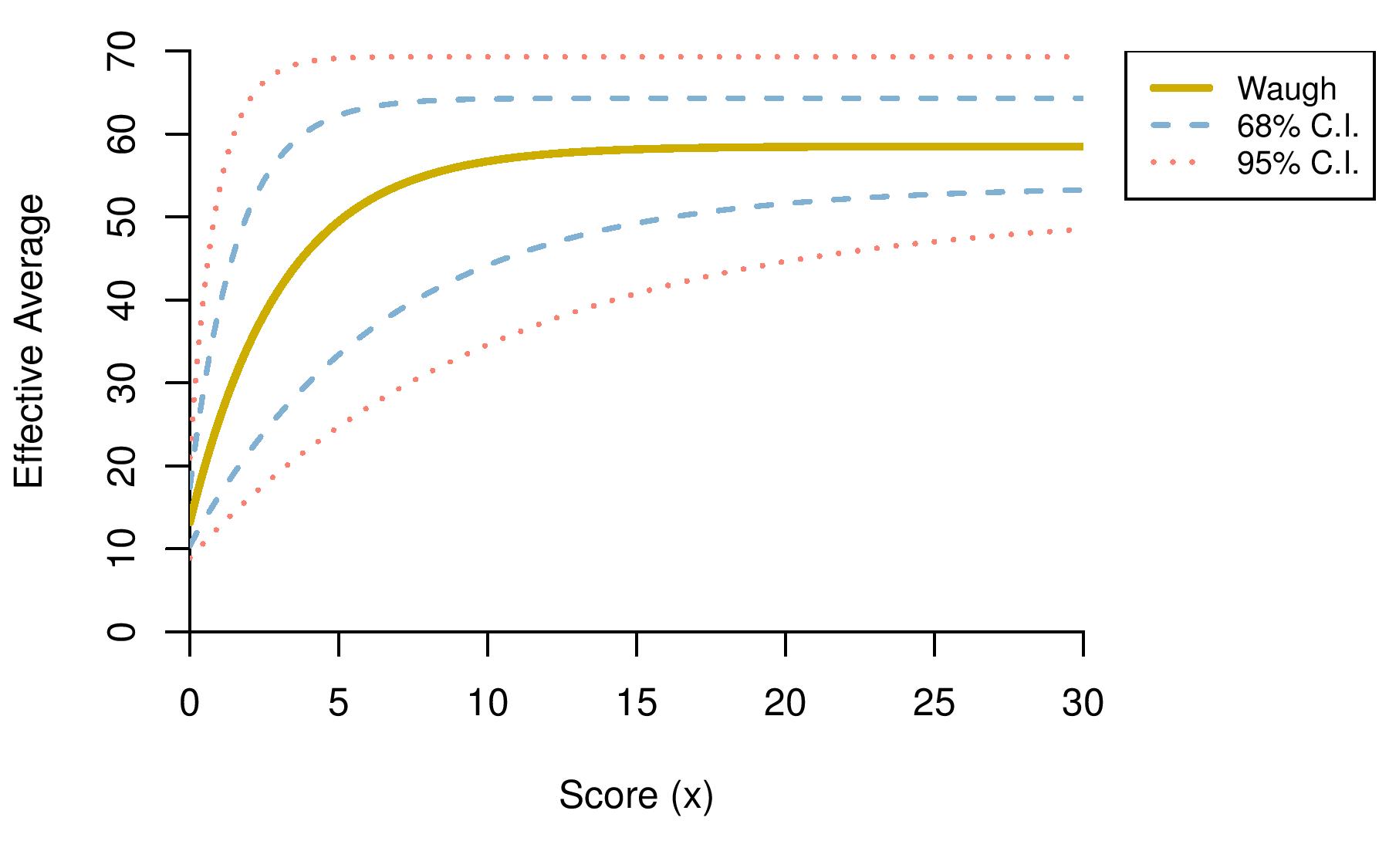}
\caption{Plot of Steve Waugh's estimated effective average $\mu(x)$, 
illustrating how his batting ability changes with his current score. 
The blue and red lines represent 68\% and 95\% credible intervals.}
\label{waugheffective}
\end{figure}

\begin{figure}[h]
\centering
  \includegraphics[width = 0.85\linewidth]{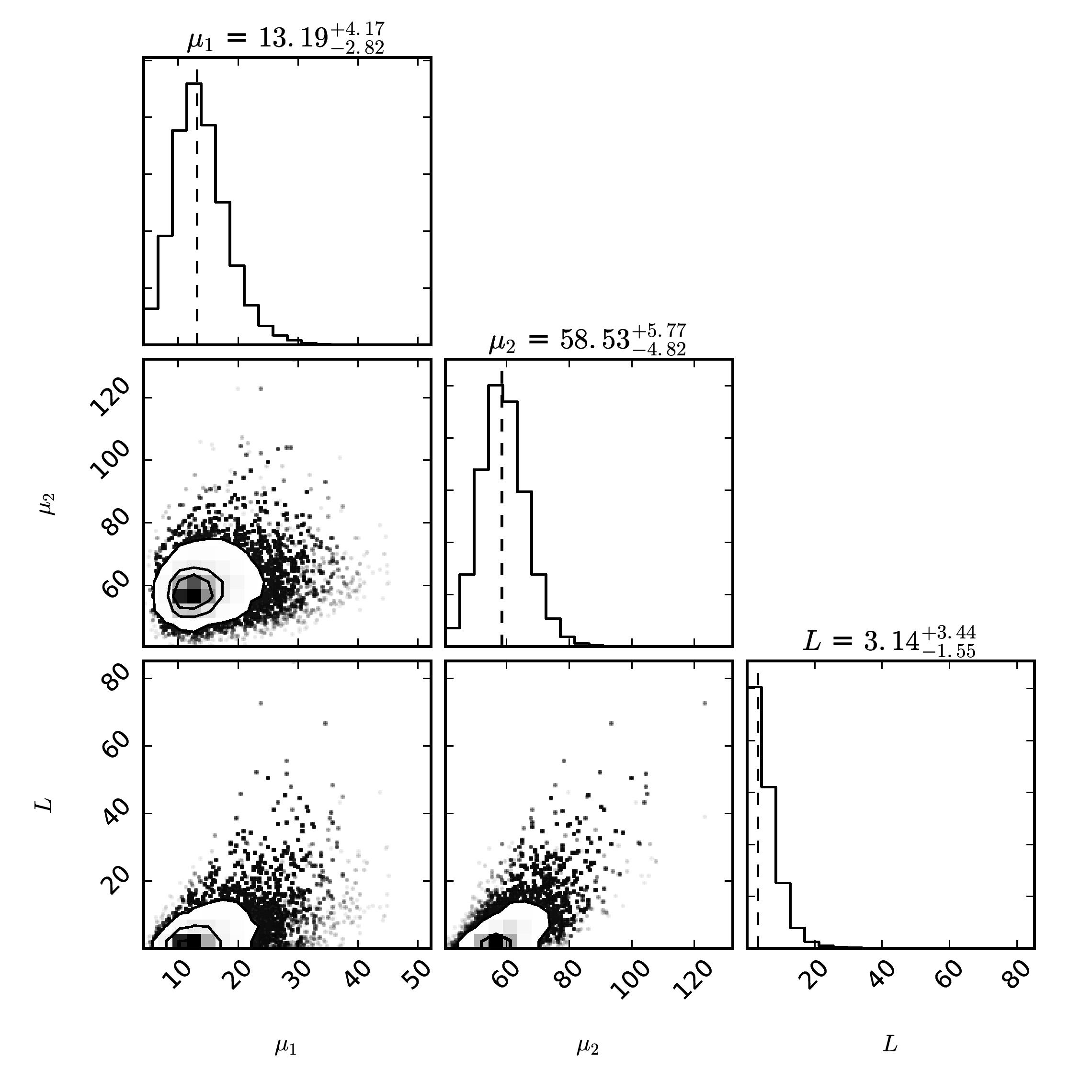}
\caption{Posterior marginal distributions for $\mu_1$, $\mu_2$ and $L$ for
Steve Waugh.
The contours represent the $50^{th}$, $68^{th}$ and $95^{th}$ percentile
limits.
Created using the corner.py package \citep{cornerpy}.}
\label{waughcorner}
\end{figure}

We drew samples from the posterior distribution for the parameters of each
player.
To illustrate the practical implications of the results, posterior samples
for former Australian captain Steve Waugh are shown in Figure
\ref{waughcorner}.
The marginal distribution for $\mu_1$ implies that Waugh arrives at the
crease batting with the ability of a player with an average of 13.2 runs.
After scoring about 3 runs, Waugh has transitioned approximately 63\% of
the way between his initial batting ability and `eye in' batting ability.
Finally, once Waugh has his eye in, he bats like a player with an average
of 58.5. Figure \ref{waugheffective} gives a visual representation of how 
well Steve Waugh is batting during the first 30 runs of his innings, including
uncertainties.

The marginal distributions in Figure \ref{waughcorner} are used to construct 
point estimates for the effective average curves (using $\mu(x; \mu_1, \mu_2, 
L)$ from Equation \ref{mux}).
These curves, seen in Figures \ref{waugheffective} and \ref{effav1},
indicate how well individual players are batting given their current score,
that is, the average number of runs they will score from a given score onwards.

\subsubsection{Posterior Summaries}

Using the marginal posterior distributions, estimates and uncertainties
were derived for the three parameters of interest for each player.
The estimates take the form posterior median $\pm$ standard deviation
and are presented in Table \ref{summtable}, together with each player's Test
career record in Table \ref{recordtable}.
We used the median as the posterior distributions are not necessarily
symmetric and some have relatively heavy tails.

Unsurprisingly, the players with the highest career averages (Brian Lara and
Steve Waugh) appear to be the best players once they have their eye in
(i.e. they have the highest $\mu_2$ estimates). 
However, it is not necessarily these players who arrive at the crease batting 
with the highest ability.
In fact, two of the players with the highest initial batting abilities,
$\mu_1$, are those with lower career Test averages, all-rounders Chris Cairns
and Shaun Pollock.
Interestingly, both players tend to bat in the middle to lower order and
have lower estimates for $\mu_2$, their `eye in' batting ability,
suggesting they don't quite have the same batting potential as the other top
order batsmen.
This outcome may be due to initial batting conditions tending to be more
difficult for batsmen in the top order, compared with those in the middle
and lower order.
Additionally, the result may derive from the aggressive nature in which
Cairns and Pollock play, meaning even when they are dismissed early they often
return to the pavilion with some runs to their name.
This gives rise to the notion that perhaps a player's strike rate (runs
scored per 100 balls faced) and batting position may be influential on the
parameter $C$ (the size of $\mu_1$ with respect to $\mu_2$).

The marginal likelihood or `evidence' was also measured for each player
analyzed using the individual player model.
In a Bayesian inference problem with parameters $\theta$ and data $d$, the
marginal likelihood of a model $M$ is the probability of the data given the
model, i.e.,
\begin{align}
Z &= 
p(d | M) = \int p(\theta | M)p(d | \theta, M) \, d\theta,
\end{align}
and is used as an input to the posterior probability of model $M$ compared to
an alternative.
Nested sampling allows us to easily calculate $Z$ \citep{skilling2006}.
In this case we can use the evidence to compare the support for our
varying-hazard model ($Z$), against a constant hazard model ($Z_0$) which has a
lognormal(20, 0.75$^2$) prior on its constant effective average $\mu$.
The logarithm of the Bayes factor between these two models is included in
Table \ref{summtable} and suggests the varying-hazard model is favoured for
all players.
As the nested sampling method used is an MCMC process, these results are
not exact, however the algorithm was run with a large number of particles
and MCMC iterations and therefore the Monte-Carlo related errors are 
negligible.

\begin{table}[h]
\caption{Test career records for analyzed players.}
\renewcommand{\arraystretch}{1.25}
{\resizebox{\textwidth}{!}
{\begin{tabular}{l c c c c c c c c c}
\hline
  Player & Matches & Innings & Not-Outs & Runs & High-Score & Average & Strike Rate & 100s & 50s \\
  \hline
  C. Cairns (NZ) & 62 & 104 & 5 & 3320 & 158 & 33.53 & 57.09 & 5 & 22 \\
  N. Hussain (ENG) & 96 & 171 & 16 & 5764 & 207 & 37.18 & 40.38 & 14 & 33 \\
  G. Kirsten (SA) & 101 & 176 & 15 & 7289 & 275 & 45.27 & 43.43 & 21 & 24\\
  J. Langer (AUS) & 105 & 182 & 12 & 7696 & 250 & 45.27 & 54.22 & 23 & 30 \\
  B. Lara (WI) & 131 & 232 & 6 & 11953 & 400* & 52.88 & 60.51 & 34 & 48 \\
  S. Pollock (SA) & 108 & 156 & 39 & 3781 & 111 & 32.31 & 52.52 & 2 & 16 \\
  S. Warne (AUS) & 145 & 199 & 17 & 3154 & 99 & 17.32 & 57.65 & 0 & 12 \\
  S. Waugh (AUS) & 168 & 260 & 46 & 10927 & 200 & 51.06 & 48.64 & 32 & 50 \\
\end{tabular}}}
\label{recordtable}
\end{table}

\begin{table}[h]
\caption{Parameter estimates and uncertainties using the individual player model.}
\renewcommand{\arraystretch}{1.25}%
{\resizebox{\textwidth}{!}
{\begin{tabular}{l c c c c c c | c r}
\hline
Player & $\mu_1$ & 68\% C.I. & $\mu_2$ & 68\% C.I. & $L$ & 68\% C.I. & log$_e$(Z) & log$_e$(Z/Z$_0$) \\
\hline
C. Cairns & $16.6 ^{+ 6.4}_{- 5.2}$ & [11.4, 23.0] & $36.1 ^{+ 4.4}_{- 3.8}$ & [32.3, 40.5] & $2.3 ^{+ 4.6}_{- 1.7}$ & [0.6, 6.9] & $-449.69$ & 1.04  \\
N. Hussain & $12.8 ^{+ 4.6}_{- 3.2}$ & [9.6, 17.4] & $40.8 ^{+ 4.3}_{- 3.5}$ & [37.3, 45.1] & $1.9 ^{+ 2.7}_{- 1.2}$ & [0.7, 4.6] & $-714.52$ & 5.88 \\
G. Kirsten & $14.4 ^{+ 4.8}_{- 3.4}$ & [11.0, 19.2] & $53.9 ^{+ 6.6}_{- 5.3}$ & [48.6, 60.5] & $6.3 ^{+ 4.6}_{- 2.9}$ & [3.4, 10.9] & $-769.79$ & 10.00 \\
J. Langer & $18.0 ^{+ 7.5}_{- 4.8}$ & [13.2, 25.5] & $49.2^{+ 5.0}_{- 4.2}$ & [45.0, 54.2] & $2.7 ^{+ 4.1}_{- 1.8}$ & [0.9, 6.8] & $-800.83$ & 3.95 \\
B. Lara & $15.1 ^{+ 4.6}_{- 3.5}$ & [11.6, 19.7] & $61.8 ^{+ 5.7}_{- 5.1}$ & [56.7, 67.5] & $6.1 ^{+ 3.9}_{- 2.7}$ & [3.4, 10.0] & $-1114.95$ & 13.37 \\
S. Pollock & $18.2 ^{+ 4.8}_{- 4.3}$ & [13.9, 23.0] & $37.4 ^{+ 5.8}_{- 4.4}$ & [33.0, 43.2] & $5.6 ^{+ 6.0}_{- 3.5}$ & [2.1, 11.6] & $-526.15$ & 1.83 \\
S. Warne & $5.3 ^{+ 1.2}_{- 0.9}$ & [4.4, 6.5] & $21.2 ^{+ 2.1}_{- 1.9}$ & [19.3, 23.3] & $1.3 ^{+ 1.2}_{- 0.8}$ & [0.5, 2.5] & $-679.77$ & 15.60 \\
S. Waugh & $13.2 ^{+ 4.2}_{- 2.8}$ & [10.4, 17.4] & $58.5 ^{+ 5.8}_{- 4.8}$ & [53.7, 64.3] & $3.1 ^{+ 3.4}_{- 1.6}$ & [1.5, 6.5] & $-1032.36$ & 13.98 \\
\hline
\textbf{Prior} & $ 6.6^{+ 12.8}_{- 5.0}$ & [1.6, 19.4] & $25.0 ^{+ 27.7}_{- 13.1}$ & [11.9, 52.7] & $ 3.0^{+ 6.7}_{- 2.3}$ & [0.7, 9.7] & N/A & N/A
\end{tabular}}}
\label{summtable}
\end{table}

These results are relatively consistent with \cite{brewer2008} (who used a
different model for $\mu(x)$); Cairns, Langer and Pollock are the best batsmen
when first arriving at the crease, and Lara and Waugh have the highest `eye in'
batting abilities.
The actual point estimates were similar in most cases, though the present model
has less uncertainty in values of $\mu_1$ (probably since $\mu_1 = \mu(0)$ in 
our model), but more uncertainty in $\mu_2$ values.
It is difficult to directly compare the transition variable $L$, as
\cite{brewer2008} used two variables to capture the change between the two
batting states.

\subsubsection{Effective Average Curves}

The posterior summaries allow us to construct a predictive hazard function
for a player's next innings, which is a slightly different point estimate 
for $\mu(x)$ than the posterior mean or median.
These are obtained by calculating the posterior predictive distribution for
a player's `next' score given the data, and deriving the hazard function
$H(x)$ corresponding to the predictive distribution using Equation~\ref{hx},
in terms of the effective average (Equation \ref{mux}).
For clarity, only the functions for four of the recognised batsman in the
analysis were included (Gary Kirsten, Justin Langer, Brian Lara and Steve Waugh).

\begin{figure}[h]
\centering
  \includegraphics[width = 0.85\linewidth]{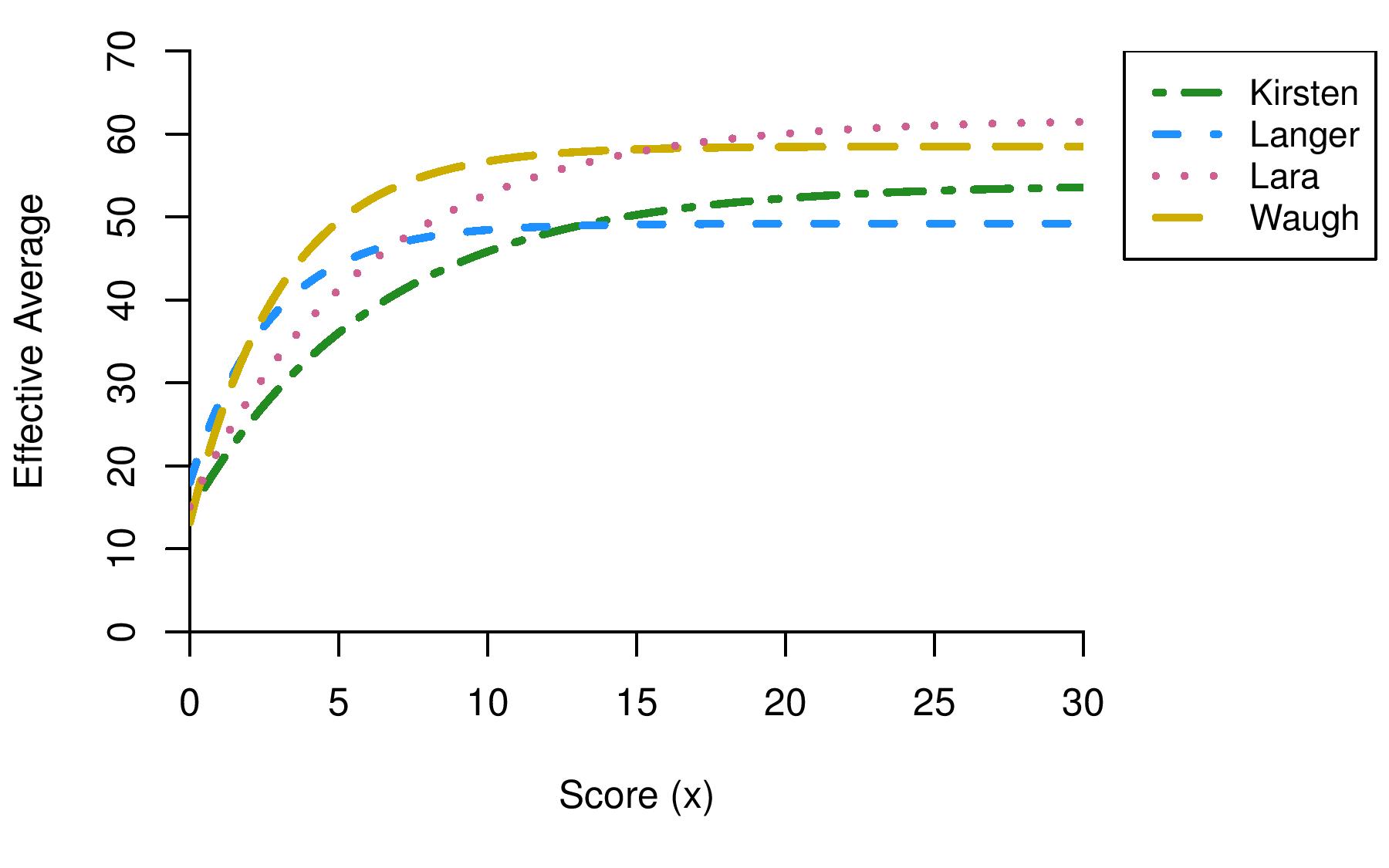}
\caption{Predictive hazard functions (a kind of point estimate for $\mu(x))$
in terms of effective average for Kirsten, Langer, Lara and Waugh.}
\label{effav1}
\end{figure}

Figure \ref{effav1} gives a visual representation of the posterior summaries
in Table \ref{summtable}.
Of the four players shown, Waugh has the lowest effective average when
first arriving at the crease.
However, Waugh gets his eye in relatively quickly and appears to be batting
better than the others after scoring just a couple of runs.
Not until scoring approximately 15 runs does Lara overtake Waugh, suggesting 
Lara is a better batsman when set at the crease
($P(\mu_{2 \ Lara} > \mu_{2 \ Waugh} | d) = 0.66$),
but takes longer to get his eye in 
($P(L_{Lara} > L_{Waugh} | d) = 0.75$).

An interesting comparison can also be made between Kirsten and Langer,
two opening batsmen with identical career Test batting averages of 47.27.
Langer arrives at the crease with a higher initial batting ability than
Kirsten ($P(\mu_{1 \ Langer} > \mu_{1 \ Kirsten} | d) = 0.70$) and is also
quicker to get his eye in ($P(L_{Langer} < L_{Kirsten} | d) = 0.77$).
Only after scoring about 13 runs, does Kirsten look to be playing better
than Langer in terms of batting ability.
This arguably makes Langer the preferred choice for an opening batsmen as it
suggests he is less susceptible at the beginning of his innings and is more
likely to succeed in his job as an opener, seeing off the new ball and
opening bowlers.
However, to come to a more substantive conclusion as to which player is
more suited to the opening role, it may also be worthwhile considering
the variables `balls faced' or `minutes batted' instead of runs scored,
especially since more traditional opening batsmen (such as Kirsten) are
known for their tendencies to score at a slower rate than other batsmen.

Due to our restriction that the hazard function must be monotonically 
decreasing (and therefore the effective average is monotonically increasing),
our estimates of a player's ability are not as erratic as those in 
\cite{kimber1993} and \cite{cai2002}.
However, it is entirely plausible, if not probable, that effects between
certain scores and a player's effective average exist.
For example, it is not uncommon to see batsmen lose concentration after
batting for a long period of time or, before or after passing a significant
milestone (e.g. scoring 50, 100, 200).
Investigating the relationship between certain scores and the effective
average warrants closer attention in future research.

\section{Hierarchical Model}

While knowing the performance of individual players is useful, we can
generalise our inference to a wider group of players by implementing a
hierarchical model structure.
Instead of applying the prior $\mu_2 \sim \textnormal{lognormal}(25, 0.75^2)$
to each player, we define hyperparameters $\nu, \sigma$ such that the prior
for each player's $\mu_2$ is

\begin{align}
\mu_{2, i} | \nu, \sigma &\sim \textnormal{LogNormal}(\nu, \sigma^2).
\end{align}

When we infer $\nu$ and $\sigma$ from the data for a group of players,
we can quantify the typical $\mu_2$ value the players are clustered around
using $\nu$, while $\sigma$ describes how much $\mu_2$ varies from player
to player.

To apply the hierarchical model, we first analyzed each player in the group 
of interest using the individual player model in Section \ref{CommonPrior}, 
and post-processed the results to reconstruct what the hierarchical 
model would have produced.
This is a common technique for calculating the output of a hierarchical 
model without having to analyze the data for all players jointly. 
\citet{hastings1970} suggested using MCMC samples for this purpose.
For an example of the same technique applied in astronomy, see 
\citet{brewer2014hierarchical}.

\subsection{Prior Distributions}

The hierarchical model is implemented by writing the prior for $\mu_2$
conditional on hyperparameters $\alpha = (\nu, \sigma)$, as
lognormal($\nu$, $\sigma^2$); rather than using a common lognormal(25, 0.75$^2$)
prior for all players.
The idea is to gain an understanding of the posterior distributions for $\nu$
and $\sigma$, rather than $\mu_2$ directly.
Whereas informal prior knowledge of cricket was used to assign the
original lognormal(25, 0.75$^2$) prior, the hierarchical model does this
more explicitly, as the prior for a player's parameters is informed by the
data from other players.
The priors over parameters $C$ and $D$ were kept constant.

We assigned flat, uninformative, uniform(1, 100) and uniform(0, 10) priors
for the hyperparameters $\alpha = (\nu, \sigma)$ respectively.
The full model specification is therefore

\begin{align}
\nu &\sim \textnormal{Uniform}(1, 100)\\
\sigma &\sim \textnormal{Uniform}(0, 10)\\
\mu_{2, i} | \nu, \sigma &\sim \textnormal{LogNormal}(\nu, \sigma^2) \\
C_i &\sim \textup{Beta}(1, 2) \\
D_i &\sim \textup{Beta}(1, 5) \\
\textnormal{log-likelihood} &\sim
    \sum_i\textnormal{(Equation~\ref{loglik})}
\end{align}
where subscript $i$ denotes the $i^{th}$ player in our sample.

The marginal posterior distribution for the hyperparameters given all of the 
data may be written in terms of expectations over the individual players' 
posterior distributions computed as in Section \ref{CommonPrior} 
\citep[see e.g.,][]{brewer2014hierarchical}

\begin{equation}
  p(\nu, \sigma | \{D_i\}) \propto p(\nu, \sigma)
\prod_{i = 1}^N 
\mathbb{E} \left[\frac{f(\mu_{2, i} | \nu, \sigma)}{\pi(\mu_{2, i})}\right]
\label{hierarchyequation}
\end{equation}
where $f(\mu_{2, i} | \nu, \sigma)$ is the lognormal$(\nu, \sigma^2)$ prior
applied to $\mu_2$ for the $i$th player, and $\pi(\mu_{2, i})$ is the
lognormal$(25, 0.75^2)$ prior that was actually used to calculate the 
posterior for each individual player. 
The expectation (i.e., each term inside the product) can be approximated 
by averaging over the posterior samples for that player.

\subsection{Data}

The data used with the hierarchical model again come from Statsguru on the
Cricinfo website. We decided the results of the model would be particularly
useful when applied to opening batsmen, as it can pinpoint players
who are susceptible at the beginning of their innings.

Given the authors' nationality and country of residence respectively, the 
hierarchical model was used to make generalized inference about opening 
batsman who have represented New Zealand.
As opening has been a position of concern for the national team for some time,
all opening batsman to play for New Zealand since 1990 (who have since retired)
were included in the study.
Any player who opened the batting for New Zealand for at least 50\% of their
innings was deemed an opening batsman, and included in the hierarchical analysis.

In most conditions, the first overs of a team's innings are considered the
most difficult to face, as the ball is new and batsmen are not used to the pace
and bounce of the pitch.
For the purposes of this discussion, in order to counter these difficult
batting conditions, we would hope that an opening batsman begins their innings
batting closer to their peak ability than a middle or lower order batsman.
That is, we might want opening batsmen to be more `robust' in the sense that 
the difference between their abilities when fresh and when set is smaller.
As such, these results can help coaches and selectors compare and identify 
players who are more or less suited to opening the batting.

\subsection{Results}
New Zealand opening batsmen were analyzed separately using the individual
player model from Section \ref{CommonPrior}.
Posterior summaries were generated for each player and are presented in order
of Test debut in Table \ref{nztable} (see Appendix~\ref{sec:nz_openers}).
Due to several players appearing in just a handful of matches, some
uncertainties are fairly large.

\begin{figure}[h]
\centering
  \includegraphics[width = \linewidth]{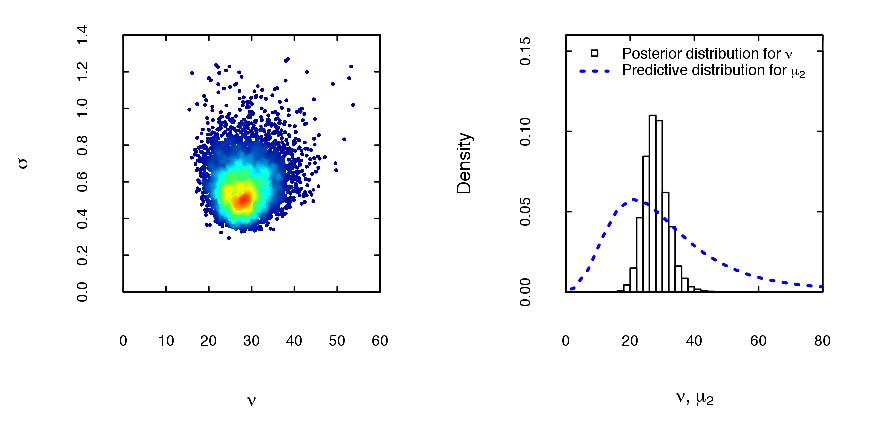}
  \caption{(Left) Joint posterior distribution for $\nu$ and $\sigma$.
  (Right) Marginal posterior distribution for $\nu$ with predictive distribution
  for $\mu_2$ overlaid.}
  \label{jointmarginal}
\end{figure}

The posterior samples for each player were then combined, allowing us to make 
posterior inferences regarding hyperparameters $\nu$ and $\sigma$, using the 
result from Equation \ref{hierarchyequation}.
The joint posterior distribution for $\nu$ and $\sigma$ is shown in Figure
\ref{jointmarginal} and represents just a small proportion of the area
covered by the uniform prior distributions, suggesting the data contained a
lot of information about the hyperparameters.

The marginal posterior distribution for $\nu$ is also shown in Figure
\ref{jointmarginal}, with the posterior predictive distribution for $\mu_2$
overlaid. Our inference regarding $\nu$ can be summarised as:
$\nu = 27.85 ^{+ 3.74}_{- 3.55}$, while $\sigma$ can be summarised as:
$\sigma = 0.54 ^{+ 0.11}_{- 0.08}$.
These results suggest our subjectively assigned lognormal$(25, 0.75^2)$ prior 
in Section~\ref{CommonPrior} was reasonably close to the actual frequency 
distribution of $\mu_2$ values among this subset of Test cricketers.

Using the hyperparameter posterior samples, we are also able to make an
informed prediction regarding the batting abilities of the `next' opening
batsman to debut for New Zealand.
Our estimates for the next opener are $\mu_1 = 9.6 ^{+ 11.7}_{- 5.7}$, 
$\mu_2 = 27.7 ^{+ 21.0}_{- 11.9}$ and $L = 3.1 ^{+ 6.0}_{- 2.4}$.
This prediction is summarized in the final row of Table \ref{nztable}.

The opening batsmen included in this study accounted for 546 separate Test 
innings. Given this moderate sample size, the uncertainties are somewhat large,
although with more data we would expect more precise inferences and
predictions.

Of course, this prediction must be taken with a grain of salt, as the New Zealand
cricketing landscape has changed drastically since the 1990s.
The ever-increasing amount of money invested in the game allows modern day
players to focus on being full-time cricketers.
The structure of the domestic cricket scene has also improved, including 
better player scouting and coaching, resulting in the best local talents being 
nurtured from a young age.
It is also worth noting we have chosen to exclude players who are 
currently playing for New Zealand.
The individual player model would likely find the current opening batsmen to
be `better' than the point estimates in Table \ref{nztable}.
As a result, our prediction is likely an underestimate of the next opening 
batsman's abilities.
Nevertheless, the prediction does highlight the difficulties New Zealand
has had in the opening position. Few batsman with an `eye in' average, let
alone a career average, of 27.7, would make many international sides on
batting ability alone.

\begin{figure}[H]
\centering
  \includegraphics[width = \linewidth]{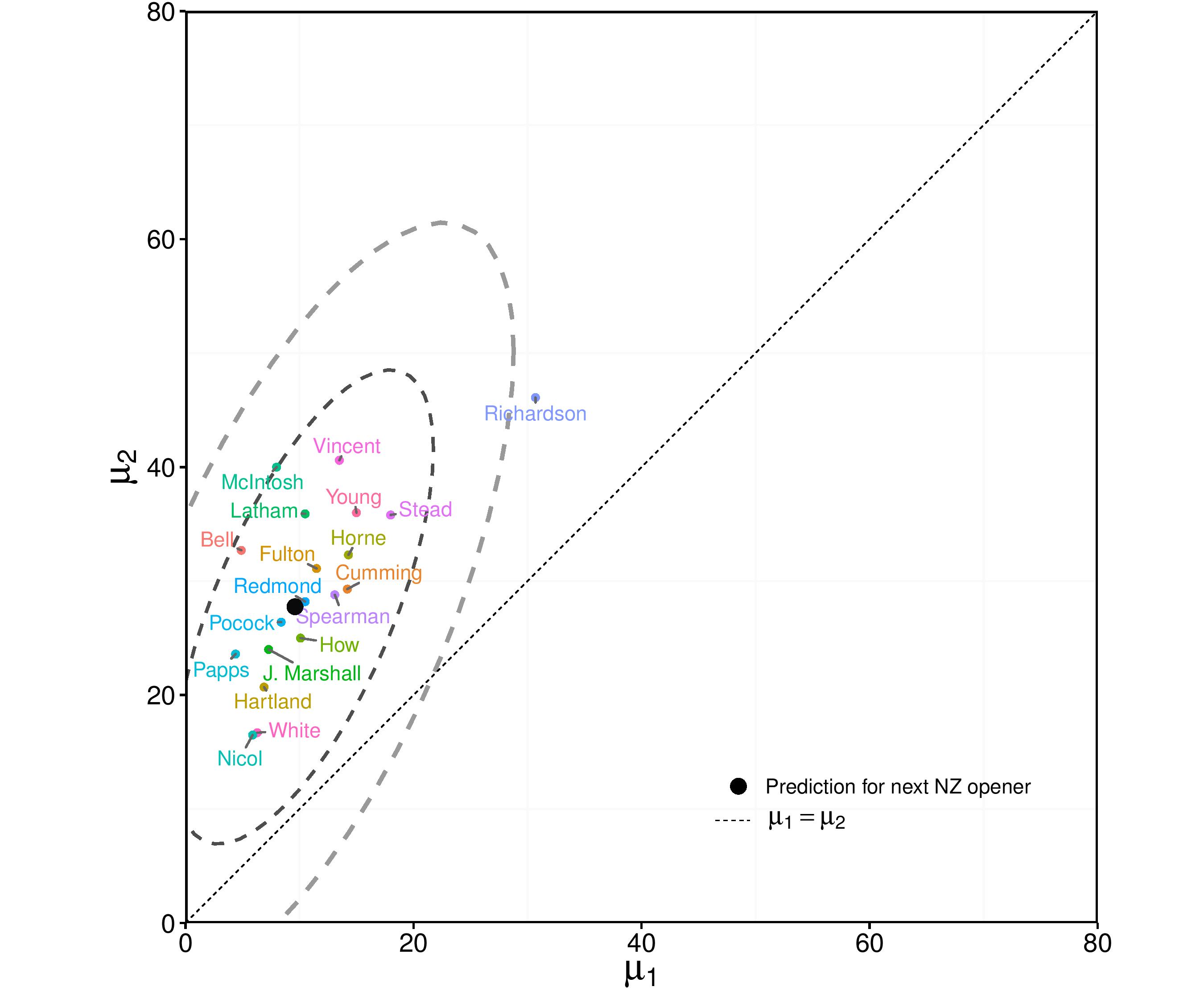}
\caption{Point estimates for all analyzed batsmen on the $\mu_1$ -- $\mu_2$ plane. 
The prediction for the next New Zealand opening batsmen is represented by the 
black dot, including 68\% (inner) and 95\% (outer) credible intervals 
(dotted ellipses).}
\label{nzopener}
\end{figure}

Figure \ref{nzopener} depicts the point estimates on the $\mu_1$ -- $\mu_2$
plane for all New Zealand openers analyzed in the study. All players fall 
within the 68\% and 95\% credible intervals of the prediction for the next 
opener, with the exception of Mark Richardson. 
Unsurprisingly, this suggests almost all players analyzed are typical of 
New Zealand opening batsmen.

Since his debut in 2001, Richardson has widely been considered New 
Zealand's only world class Test opener to play in the current millennium.
Figure \ref{nzopener} certainly suggests Richardson is class apart from his
compatriots, as he is the only player to fall outside the 95\% credible 
interval.
Estimates for both Richardson's initial and `eye in' abilities are also 
considerably higher than the predicted abilities for the next opening batsman: 
$P(\mu_{1 \ Richardson} > \mu_{1 \ Predicted}) = 0.91$, and 
$P(\mu_{2 \ Richardson} > \mu_{2 \ Predicted}) = 0.81$.
If our notion that opening batsmen should be more `robust' than middle order 
batsmen, then Richardson certainly appears to be the ideal opener given
his very high initial batting ability.

\section{Conclusions}
This paper has presented a Bayesian approach to modelling the hazard function
for batsmen in Test cricket.
The results support common cricketing beliefs, that batsmen are more
vulnerable at the beginning of their innings and improve as they score
more runs.
However, the model provides the added advantage of quantifying the effect
and its significance for individual players.
Interestingly, career average doesn't necessarily correlate to a higher
initial batting ability, instead this may be more directly related to other
factors such as batting position and strike rate.
The speed of transition between batting states gives an indication of how
long a player needs to be batting at their best, although the only measure
of this was the number of runs scored.
In the future, also including variables such as the number of balls faced
and minutes at the crease may help better identify the speed of a player's 
transition.

Applying the model to a wider group of players allows us to make an
informed prediction about the abilities of next opening batsman to debut
for New Zealand.
However, our concept of openers being more `robust' than other players isn't
widely supported among the openers analyzed, although this may be due to the
talent pool we focussed on.
Using the hierarchical model for a country that has produced a number of
world class openers in the recent past (e.g. Australia, England, India,
South Africa), may yield a different conclusion.

The simple, varying-hazard model provides us with a foundation upon which we
can add further parameters for future research.
In the present paper we have assumed conditional independence between innings 
across a batsman's career.
In reality, it is far more likely that some time dependent effect exists
between parameters $\mu_1$, $\mu_2$ and $L$.
Temporal variation may exist on two planes, long-term changes due to factors
such as age and experience and short-term changes due to player form and
confidence.
Allowing for more complex relationships between our parameters of interest
will give our model the ability to answer more difficult questions, such as
how long it takes for a new Test batsman to find their feet on the
international scene and start performing at their best.

It may also be worthwhile to explore the relationship between players'
first-class and Test batting averages, identifying characteristics that
translate to smaller and larger discrepancies between the two averages.
Another study may take the present analysis further, looking more closely at 
how well players are batting during their innings, to identify trends in
the hazard function near certain scores.
Given the statistical nature of the game, milestones can play a large role
in a batsman's innings, possibly causing lapses in judgement when
nearing significant scores (e.g. 50, 100).
Undertaking a deeper analysis may allow us to tentatively confirm or
refute popular cricketing superstitions, such as the commentator's 
favourite, the `nervous nineties'.

\section*{Acknowledgements}
It is a pleasure to thank Berian James (Square), Mathew Varidel (Sydney), 
Ewan Cameron (Oxford), Ben Stevenson (St Andrews), Matt Francis (Ambiata)
and Thomas Lumley (Auckland) for their helpful discussions.
We would also like to thank the reviewers and journal editors for their useful
comments and suggestions.

\bibliographystyle{DeGruyter}
\bibliography{references}

\appendix
\section{New Zealand Opening Batsmen}\label{sec:nz_openers}

\begin{table}[H]
\caption{Test career records for New Zealand opening batsmen who debuted since
1990.}
\renewcommand{\arraystretch}{1.25}
{\resizebox{\textwidth}{!}
{\begin{tabular}{l c c c c c c c c c}
\hline
  Player & Matches & Innings & Not-Outs & Runs & High-Score & Average & Strike Rate & 100s & 50s \\
  \hline
  D. White & 2 & 4 & 0 & 31 & 18 & 7.75 & 33.33 & 0 & 0 \\
  B. Hartland & 9 & 18 & 0 & 303 & 52 & 16.83 & 31.33 & 0 & 1 \\
  R. Latham & 4 & 7 & 0 & 219 & 119 & 31.28 & 48.99 & 1 & 0 \\
  B. Pocock & 15 & 29 & 0 & 665 & 85 & 22.93 & 29.80 & 0 & 6 \\
  B. Young & 35 & 68 & 4 & 2034 & 267* & 31.78 & 38.95 & 2 & 12 \\
  C. Spearman & 19 & 37 & 2 & 922 & 112 & 26.34 & 41.68 & 1 & 3 \\
  M. Horne & 35 & 65 & 2 & 1788 & 157 & 28.38 & 40.78 & 4 & 5 \\
  M. Bell & 18 & 32 & 2 & 729 & 107 & 24.30 & 37.81 & 2 & 3 \\
  G. Stead & 5 & 8 & 0 & 278 & 78 & 34.75 & 41.43 & 0 & 2 \\
  M. Richardson & 38 & 65 & 3 & 2776 & 145 & 44.77 & 37.66 & 4 & 19 \\
  L. Vincent & 23 & 40 & 1 & 1332 & 224 & 34.15 & 47.11 & 3 & 9 \\
  M. Papps & 8 & 16 & 1 & 246 & 86 & 16.40 & 35.34 & 0 & 2 \\
  C. Cumming & 11 & 19 & 2 & 441 & 74 & 25.94 & 34.86 & 0 & 1 \\
  J. Marshall & 7 & 11 & 0 & 218 & 52 & 19.81 & 39.06 & 0 & 1 \\
  P. Fulton & 23 & 39 & 1 & 967 & 136 & 25.44 & 39.27 & 2 & 5 \\
  J. How & 19 & 35 & 1 & 772 & 92 & 22.70 & 50.45 & 0 & 4 \\
  A. Redmond & 8 & 16 & 1 & 325 & 83 & 21.66 & 39.01 & 0 & 2 \\
  T. McIntosh & 17 & 33 & 2 & 854 & 136 & 27.54 & 36.20 & 2 & 4 \\
  R. Nicol & 2 & 4 & 0 & 28 & 19 & 7.00 & 26.66 & 0 & 0 \\
\end{tabular}}}
\label{NZrecordtable}
\end{table}

\begin{table}[H]
\caption{Posterior summaries for all New Zealand opening batsmen since 1990, 
including our estimate for the next opener to debut for New Zealand. 
Players are ordered by Test debut date (older to recent).}
\renewcommand{\arraystretch}{1.35}%
{\resizebox{\textwidth}{!}
{\begin{tabular}{l c c c c c c c c c}
\hline
Player & $\mu_1$ & 68\% C.I. & $\mu_2$ & 68\% C.I. & $L$ & 68\% C.I. \\
\hline
D. White & $6.3 ^{+ 5.6}_{- 3.3}$ & [3.0, 11.9] & $16.7 ^{+ 13.9}_{- 7.0}$ & [9.7, 30.6] & $2.7 ^{+ 4.9}_{- 2.1}$ & [0.6, 7.6] \\
B. Hartland & $6.9 ^{+ 4.6}_{- 2.9}$ & [4.0, 11.5] & $20.7 ^{+ 6.6}_{- 4.6}$ & [16.1, 27.3] & $1.9 ^{+ 3.3}_{- 1.4}$ & [0.5, 5.2] \\ 
R. Latham & $10.5 ^{+ 9.9}_{- 5.8}$ & [4.7, 24.4] & $35.9 ^{+ 17.5}_{- 10.6}$ & [25.3, 53.4] & $4.1 ^{+ 7.6}_{- 3.2}$ & [0.9, 11.7] \\ 
B. Pocock & $8.4 ^{+ 5.5}_{- 3.3}$ & [5.1, 13.9] & $26.4 ^{+ 6.4}_{- 4.7}$ & [21.7, 32.8] & $1.9 ^{+ 3.4}_{- 1.4}$ & [0.5, 5.3] \\ 
B. Young & $15.0 ^{+ 5.9}_{- 4.9}$ & [10.1 , 20.1] & $36.0 ^{+ 6.4}_{- 4.8}$ & [31.2, 42.4] & $4.4 ^{+ 6.3}_{- 3.0}$ & [1.4, 10.7] \\ 
C. Spearman & $13.1 ^{+ 6.2}_{- 4.8}$ & [8.3, 19.3] & $28.8 ^{+ 5.9}_{- 4.8}$ & [24.0, 34.7] & $2.0 ^{+ 3.4}_{- 1.5}$ & [0.5, 5.4] \\ 
M. Horne & $14.3 ^{+ 5.4}_{- 4.3}$ & [10.0, 19.7] & $32.3 ^{+ 5.7}_{- 4.5}$ & [27.8, 38.0] & $4.4 ^{+ 5.1}_{- 2.7}$ & [1.7, 9.5] \\ 
M. Bell & $4.9 ^{+ 3.0}_{- 1.8}$ & [3.1, 7.9] & $32.7 ^{+ 10.1}_{- 6.7}$ & [26.0, 42.8] & $3.2 ^{+ 5.4}_{- 2.5}$ & [0.7, 8.6] \\ 
G. Stead & $18.0 ^{+ 11.5}_{- 8.6}$ & [9.4, 29.5] & $35.8 ^{+ 14.9}_{- 9.8}$ & [26.0, 50.7] & $3.1 ^{+ 6.7}_{- 2.4}$ & [0.7, 9.8] \\ 
M. Richardson & $30.7 ^{+ 8.5}_{- 8.8}$ & [21.9, 39.2] & $46.1 ^{+ 6.8}_{- 5.6}$ & [40.5, 52.9] & $3.6 ^{+ 6.9}_{- 2.8}$ & [0.8, 10.5] \\ 
L. Vincent & $13.5 ^{+ 7.0}_{- 5.2}$ & [8.3, 20.5] & $40.6 ^{+ 10.1}_{- 7.4}$ & [33.2, 50.7] & $6.0 ^{+ 7.7}_{- 4.5}$ & [1.5, 13.7] \\ 
M. Papps & $4.4 ^{+ 3.3}_{- 1.9}$ & [2.5, 7.7] & $23.6 ^{+ 11.0}_{- 6.2}$ & [17.4, 34.6] & $3.0 ^{+ 5.1}_{- 2.2}$ & [0.8, 8.1] \\ 
C. Cumming & $14.2 ^{+ 7.0}_{- 5.7}$ & [8.5, 21.2] & $29.3 ^{+ 9.3}_{- 6.5}$ & [22.8, 38.6] & $3.3 ^{+ 5.9}_{- 2.5}$ & [0.8, 9.2] \\ 
J. Marshall & $7.3 ^{+ 5.9}_{- 3.7}$ & [3.6, 13.2] & $24.0 ^{+ 9.8}_{- 6.3}$ & [17.7, 33.8] & $2.1 ^{+ 3.7}_{- 1.6}$ & [0.5, 5.8] \\ 
P. Fulton & $11.5 ^{+ 5.0}_{- 4.0}$ & [7.5, 16.5] & $31.1 ^{+ 8.4}_{- 5.9}$ & [25.2, 39.5] & $5.4 ^{+ 6.7}_{- 3.4}$ & [2.0, 12.1] \\ 
J. How & $10.1 ^{+ 5.6}_{- 3.9}$ & [6.2, 15.7] & $25.0 ^{+ 5.3}_{- 4.1}$ & [20.9, 30.3] & $1.2 ^{+ 2.7}_{- 0.9}$ & [0.3 3.9] \\ 
A. Redmond & $10.5 ^{+ 5.7}_{- 4.2}$ & [6.3, 16.2] & $28.2 ^{+ 11.9}_{- 7.3}$ & [20.9, 40.1] & $5.2 ^{+ 6.9}_{- 3.4}$ & [1.8, 12.1] \\ 
T. McIntosh & $8.0 ^{+ 4.5}_{- 3.0}$ & [5.0, 12.5] & $40.0 ^{+ 13.8}_{- 9.2}$ & [30.8, 53.8] & $9.2 ^{+ 8.0}_{- 5.3}$ & [3.9, 17.2] \\ 
R. Nicol & $5.9 ^{+ 5.4}_{- 3.0}$ & [2.9, 11.3] & $16.5 ^{+ 13.7}_{- 7.0}$ & [9.5, 30.2] & $3.0 ^{+ 5.0}_{- 2.2}$ & [0.8, 8.0] \\ 
\hline
\textbf{NZ Opener} & $9.6 ^{+ 11.7}_{- 5.7}$ & [4.0, 21.3] & $27.7 ^{+ 21.0}_{- 11.9}$ & [15.8, 48.7] & $3.1 ^{+ 6.0}_{- 2.4}$ & [0.8, 9.1] \\
\hline
\end{tabular}}}
\label{nztable}
\end{table}

\end{document}